\documentclass[prl,amsfonts,twocolumn]{revtex4}
\usepackage{graphicx, epsfig}
\usepackage{color}
\usepackage{mathrsfs}
\usepackage{amsmath}

\newcommand{\be}{\begin{equation}}
\newcommand{\ee}{\end{equation}}
\newcommand{\bea}{\begin{eqnarray}}
\newcommand{\eea}{\end{eqnarray}}

\newcommand{\gapp}{\mathrel{\raise.3ex\hbox{$>$}\mkern-14mu \lower0.6ex\hbox{$\sim$}}}
\newcommand{\lapp}{\mathrel{\raise.3ex\hbox{$<$}\mkern-14mu \lower0.6ex\hbox{$\sim$}}}
\def\bbox{{\,\lower0.9pt\vbox{\hrule \hbox{\vrule height 0.2 cm
\hskip 0.2 cm \vrule  height 0.2 cm}\hrule}\,}}
\newcommand{\beq}{\begin{equation}}
\newcommand{\eeq}{\end{equation}}

\begin{document}
\title{Reply to the``Comment on: Detecting Vanishing Dimensions Via Primordial Gravitational Wave Astronomy''}
\author{Jonas Mureika$^1$, and Dejan Stojkovic$^2$}
\affiliation{$^1$Department of Physics, Loyola Marymount University, Los Angeles, CA~~90045}
\affiliation{$^2$Department of Physics,
SUNY at Buffalo, Buffalo, NY 14260-1500}

\begin{abstract}
\widetext
The ``Comment on: Detecting Vanishing Dimensions Via Primordial Gravitational Wave Astronomy'' [arXiv:1104.1223] is misleading and premised on a misinterpretation of the main content of Phys. Rev. Lett. 106, 101101 (2011) [arXiv:1102.3434]. The main claim in the comment - that in some exotic theories different from general relativity (GR) there might be local degrees of freedom even in lower dimensional spaces - is trivial. Nevertheless, the authors of the comment fail to come-up with a single self-consistent example.   This claim, however, has no implications for our paper, in which we make it clear we are working within the framework of ``vanishing'' or ``evolving'' dimensions as defined in  arXiv:1003.5914.
\end{abstract}

 %%%%%%%%%%%%%%%%%%%%%%%%%%%%%%%%%%%%%%%%%%%%%%%%%%%%%%%

\maketitle
The comment written by Sotiriou {\it et al.} \cite{comment} is misleading and premised on a misinterpretation of the main content of \cite{Mureika:2011bv}. The authors suggest that in some exotic theories different from general relativity (GR) there might be local degrees of freedom even in lower dimensional spaces. This conclusion, however, has no implications for our paper, in which we make it clear we are working within the framework of ``vanishing'' or ``evolving'' dimensions as defined in \cite{dejan1}. The foundation of this approach is to keep the standard fundamental theories as they are ({\it i.e.} the standard model, GR, QCD, {\it etc...}) and change the dimensionality of the background on which they are defined.  As such, the lower dimensional theory of gravity we consider is simply GR.

In the introductory paragraph we anecdotally mention as a courtesy other models in which {\rm the effective dimensionality} of space is reduced at high energies. But these models are physically unrelated to our conclusions.  The authors of the comment seem to concentrate on this paragraph as providing the physical basis for the scenario under study.  We believe it is clearly indicated, however, the mechanism we are dealing with is
% including the abstract which clearly says what we are set to do - "We propose a robust and independent test for this new paradigm"
the vanishing dimensions scenario \cite{dejan1}.
This is outlined in a variety of locations, including the discussion immediately after the first paragraph, including all the equations in the paper, and concluding that ``This indicates that gravitational wave astronomy can be used as a tool for probing the novel `vanishing dimensions' framework.''

The Comment furthermore misinterprets not only our work, but also that of other authors by misrepresenting the dimensionality of their model (see \cite{others}).

%Already at this point, it is clear that the author's comment has nothing to do with our publication.

The main emphasis of the Comment is dedicated to the claim that models other than GR generically have local degrees of freedom in lower dimensions.  The authors discuss this in the context of Horava gravity \cite{horava}, although even in this case the authors admit that the dispersion relation would be {\it explicitly}  strongly Lorentz invariance violating.  This would almost automatically imply exclusion by strong constraints on Lorentz violating effects coming from the astrophysical observations.  On a pure theoretical basis, it has been shown that such theories are dynamically inconsistent \cite{Henneaux:2009zb}. While there are some reformulations of the original proposal in the literature which were claimed to be self-consistent, they significantly differ from the original model so that it is very unlikely they can serve as a UV completion of GR. We respectfully submit that their chosen example does not yield the `generic behavior' that it is claimed to represent.

From an experimental perspective, the observed cosmic ray alignment provides supporting evidence for the framework in question, and certainly not for the exotic gravity models mentioned by the authors of the comment.  For the event to be aligned on the plane, a real 2D propagation is needed, which is a generic feature of the model we consider \cite{dejan1}. The terms ``robust and independent'' refer to the other existing and/or proposed tests like alignment of cosmic ray showers and LHC phenomenology. While the LHC signature would depend on the exact way the scattering happens (regular lattice, random lattice...), 2D gravity waves can not exist in this context no matter how these 2D planes are oriented toward each other.

%We also find the sentence ``But the SD is not the quantity that appears in the Feynman loop
%integrals (as in the suggestions made in [1]);...'' extremely misleading, since nowhere in the text is such a claim made.

 We note the web version of our paper which appeared {\bf prior to} the Comment states explicitly in the conclusions:
"We showed that, under reasonable assumptions, this cutoff frequency may be accessible to future gravitational wave detectors such as LISA. This conclusion may change if the history of the early universe is radically different from the standard picture. Since the standard cosmology must kick in at $T \sim$ MeV (nucleosynthesis), and the dimensional cross-over happens at $T \sim$TeV, one has large freedom in formulating an underlying cosmological model of evolving dimensions." This statement leaves room for the exotic lower-dimensional theories different from GR and makes the motivation for writing the Comment unfounded.

To conclude,  we reiterate that the authors' main assertion, that in some exotic theories different from GR there might be local degrees of freedom even in lower dimensional spaces, is  not related to the conclusions drawn in \cite{Mureika:2011bv}. The Comment however has one positive aspect. It expands on our assertion made in conclusions that in possible extensions and variations of the "vanishing dimensions" scenario where the lower dimensional gravity significantly differs from GR, the dimensional cross-over will also be clearly marked by the change in nature of gravity waves, as the authors of the Comment also admit. It would be interesting to calculate the details of the gravity wave signature of the dimensional transition in these models.
%Even so, the authors of the comment fail to come up with a valid example.  Finally, regardless of this issue, the comment has very little or nothing to do with the publication that the comment is refereing to .

%
%\pagebreak
%\begin{figure}[h]
%\begin{center}
%\leavevmode
%%\includegraphics[scale=0.6]{}
%\caption{.}
%\label{fig0}
%\end{center}
%\end{figure}
 \end{document}